 \documentclass[preprint,12pt]{elsarticle}
 \usepackage{amssymb}
 \usepackage{amsmath}
 \newtheorem{theorem}{Theorem}
 
 \newtheorem{lemma}{Lemma}
 
 \newtheorem{defi}{Definition}
 \journal{Journal Name}

\begin{document}
 
 \begin{frontmatter}
 
 %% Title, authors and addresses
 
 %% use the tnoteref command within \title for footnotes;
 %% use the tnotetext command for the associated footnote;
 %% use the fnref command within \author or \address for footnotes;
 %% use the fntext command for the associated footnote;
 %% use the corref command within \author for corresponding author footnotes;
 %% use the cortext command for the associated footnote;
 %% use the ead command for the email address,
 %% and the form \ead[url] for the home page:
 %%
 %% \title{Title \tnoteref{label1}}
 %% \tnotetext[label1]{}
 %% \author{Name \corref{cor1} \fnref{label2}}
 %% \ead{email address}
 %% \ead[url]{home page}
 %% \fntext[label2]{}
 %% \cortext[cor1]{}
 %% \address{Address \fnref{label3}}
 %% \fntext[label3]{}
 
 \title{Analytical Volume Analysis for the Finite-time Controllable Region of the Linear Discrete-time Systems
 \thanks{Work supported by the National Natural Science Foundation of China (Grant No. 61273005)}}
 
 %% use optional labels to link authors explicitly to addresses:
 %% \author[label1,label2]{<author name>}
 %% \address[label1]{<address>}
 %% \address[label2]{<address>}
 
 \author{Mingwang Zhao}
 
 \address{Information Science and Engineering School, Wuhan University of Science and Technology, Wuhan, Hubei, 430081, China \\
 Tel.: +86-27-68863897 \\
 Work supported by the National Natural Science Foundation of China (Grant No. 61273005)}
 % \email{zhaomingwang@wust.edu.cn} 
 
 \begin{abstract}
 %% Text of abstract
 In this paper, the works on the analytical volume analysis for the controllable regions of the linear discrete-time (LDT) systems in papers \cite{zhaomw202001} and \cite {zhaomw202004} are discussed further and a new theorem on the analytical computing for the finite-time controllability zonotope (controllable region) of LDT systems are proven. And then, three analytical factors describing the control capability of the systems are deconstructed successfully from the analytical volume expression of the controllable region. Finally, the theorem is generalized to three cases: the narrow controllable region, the matrix $A$ with $n$ negative eigenvalues, the linear continuous-time systems. 
 \end{abstract}
 
 \begin{keyword}
 volume computation \sep control capability \sep analytical computation \sep controllable region \sep reachable region \sep zonotope \sep linear discrete-time systems
 
 %% keywords here, in the form: keyword \sep keyword
 
 %% MSC codes here, in the form: \MSC code \sep code
 %% or \MSC[2008] code \sep code (2000 is the default)
 
 \end{keyword}
 
 \end{frontmatter}
 
 %%
 %% Start line numbering here if you want
 %%
% \linenumbers
 
 %% main text
 
 \section{Introduction}
 \label{S:1}
 
 Since the state controllabilty, a concept decribing whethere the state variables and space of the dynamical systems are controlled or not by the input variables, is put forward by R. Kalman, et al, in 1960's \cite{KalmHoNar1963}, the further studies on describing quantitatively the control capability of the input variables to the state space attract the attentions of many researchers in control theory field. Based on the analysis of the controllability Gramian matrix and the mobile analysis of the system eigenvalues, some pioneering works on quantizing control capability (defined for describing the control ability and efficiency in paper \cite {zhaomw202003})are made \cite{VanCari1982}, \cite{Georges1995}, \cite{PasaZamEul2014}, and \cite{Ilkturk2015}. In recent years, more systematicness and deep-going works about that are got \cite{zhaomw202002}, \cite{zhaomw202003}, \cite{zhaomw202004}, and \cite{zhaomw202101}, such that, 
 \begin{enumerate}
 \item puting forth the definitions, computing methods and deconstruction of the control capability of the inputs to states; 
 
 \item proving relations among the control capability of the open-loop controlled plants, control Strategy Space of the controller, and the performance of the closed-loop control systems; 
 
 \item constructing relations between the control capability and the waste time$ \setminus$energe in the control process of the linear dynamcial systems.
  \end{enumerate}
 \noindent These studies reveal profoundly the dynamical systems and their properties for being controlled, and then it lays a good foundation for better analysis and design of control systems.
 
 In papers \cite{zhaomw202003} and \cite{zhaomw202004}, a kind of control capability of the input variables with the minimum control time attribute is defined and discussed. Based on the computation and optimization of the control capability, the bigger control strategy space, the better closed-loop performace and the strongger robustness of the controller can be gotten. In fact, the time-attribute control capability is the control ablity of the inputs with bounded amplitudes and then its computation and analysis can be regarded as the volume computation and analysis of a special zonotope, namely a controllable region, generated by a matrix pair $ \{A,B \}$, where the matrices $A$ and $B$ are respectively the system matrix and input matrix in the state space models of the linear systems. In paper \cite{zhaomw202001}, the definitions and the volume computation of the controllable region are discussed in detail, and recurssive equations for the volume computation of the finite-time controllable region, and then the analytical equations for the infinite-time controllable region when the eigenvalues $ \lambda_i$ of the matrix $A$ satisfy $ \left \{ \lambda_i \in(0,1), i= \overline {1,n} \right \}$ are got.
 More significantly, by deconstrucing these volume computing equations \cite{zhaomw202004}, some analytical factors describe the control capability are gotten
 and more extensive and in-depth analysis and optimization of the control capability can be carried out.

 In this paper, the analytical volume computation for the finite-time controllable zonotope (FTCZ) of the linear discrete-time (LDT) systems will be discussed further based on the results in paper \cite{zhaomw202001} \cite{zhaomw202004}, and then some analytical factors describe the control capability will be deconstructed from the analytical computing equation of the zonotope volume.

 \section {The Definition and the Some Results on the Volume Computation for the Special Zonotope}
 
 \subsection {The definition and the volume computation of the zonotope}
 
 In papers \cite{McMu1971}, \cite{GovKri2010}, \cite{BecRob2015}, and \cite{zhaomw202001}, a $n$-dimensional ($n$-D) zonotope spaned by a set of the $n$-D vectors is defined as follows.
 \begin{defi} \label{de:d22020101} The zonotope spanned by the $n$-D vectors of matrix $Z_{m}=[z_{1},z_{2}, \dots,z_{m}] \in R^{n \times m}$ and the parameter set with a finite interval is defined as
 \begin{equation} \label{eq:e220201}
 C_{q}(Z_{m}) = \left \{ \left. \sum_{i=1}^{m} c_{i}z_{i} \right| \forall c_{i} \in[0,1], i= \overline{1,m} \right \}
 \end{equation}
 \noindent where $q= \textrm{rank}(A_{m}), c_{i} \left(i=
 \overline{1,m} \right)$ are the parameters representing the
 zonotope, and vectors $z_i(i= \overline{1,m})$ are called as the generators of the zonotopes.
 \end{defi}
 
 According to the knowledge of the geometry and matrix algebra, the volume computing of the $n$-D zonotope can be deduceed as the following matrix computation.
 \begin{theorem} \label{th:t22020101}
 For any full row rank matrix $Z_{m} \in R^{n \times m}$, the volume of the $n$-D zonotope $C_{n}(Z_{m})$ spanned by the vectors of $Z_{m}$ can be computed as
 \begin{equation} \label{eq:e220202}
 V_{n} \left(C_{n}(Z_{m}) \right)= \sum_{(i_{1},i_{2}, \dots,i_{n}) \in \Omega_{1,m}^{n}} \left| \det \left[z_{i_{1}},z_{i_{2}}, \cdots,z_{i_{n}} \right] \right|
 \end{equation}
 \noindent where the $n$-tuple set $ \Omega_{m_1,m_2}^{n}$ consists of all possible $n$-tuples $(i_{1},i_{2}, \dots,i_{n})$ whose elements are picked from the set $ \left \{m_1,m_1+1, \cdots,m_2 \right \}$ and are sorted by their values. The computational complexity of the volume-computation method, \textit{i.e.}, the times computing the $n \times n$ determinant values, is
 \begin{equation} \label{eq:e220203}
 \frac{m!}{ \left(m-n \right)!n!}
 \end{equation}
 times, noted as the polynomial time $ \mathcal{O}(m^{n})$ on the vector number $m$. 
 \end{theorem}

 \subsection {The definition of the controllable region}
 
 In the dynamics analysis and control theory fields, the LDT systems can be modeled as follows
 \begin{equation}
 x_{k+1}=Ax_{k}+Bu_{k}, \quad x_{k} \in R^{n},u_{k} \in R^{r}, \label{eq:e220204}
 \end{equation}
 \noindent where $x_{k}$ and $u_{k}$ are the state variable and input
 variable, respectively, and matrices $A \in R^{n \times n}$ and $B \in
 R^{n \times r}$ are the state matrix and input matrix, respectively,
 in the system models \cite{Kailath1980}, \cite{Chen1998}. 
 
 For many practical engineering systems, the input variables $u_k$ are bounded or with the input-saturation property, and then the bounded and saturated expression can be normalized as follows
 $$ \left \Vert u_k \right \Vert _ \infty \le 1 $$
 Based on the above normalized expression, the controllable region can be defined clearly. The so-called controllable region here is a broad appellation and, in reality, can be divided into two cases, narrow controllable region and reachable region. 
 The $N$-steps narrow controllable region is the state region in which all states can be stabled to the origin in the state space by the $N$-stpes bounded inputs and it also can be called as the recover region. Correspondingly, the $N$-steps reachable region is the state region in which all states can be reached from the origin by the $N$-stpes bounded inputs. In fact, as two gemmetries, the narrow controllable region and the reachable region can be converted to each other through linear transformation \cite{zhaomw202004}. Therefore, the reachablity region as a broad controllable region is discuss in detail and the obtained resulte can be generlized to the narrow controllable region.

 For the state controllabilty analysis, the $N$-steps controllable region of the LDT systems, i.e., the state space generated by the $N$-steps bounded inputs $U_N= \left [u_0^T,u_1^T, \dots ,u_{N-1}^T \right]^T$ can be described as follows
 \begin{defi} \label{de:d22020102} The $N$-steps controllable region generated by the matrix pair $ \{A,B \}$ in the LDT models and the parameter set with a finite interval are defined as
 \begin{align}
 R_N^d(A) & = \left \{ \sum_{k=0} ^{N-1} A^k B u_k, \quad \left \Vert u_k \right \Vert _ \infty \le1 \right \} \notag \\
 & = \left \{ \sum_{i=1}^{rN} c_{i}p_{i} , \quad \forall c_{i} \in[0,1],i= \overline{1,rN} \right \} \label{eq:e220205}
 \end{align}
 \noindent where $P_{N} ^d= \left[B,AB, \cdots,A^{N-1}B \right]= \left[p_1,p_2, \cdots,p_{rN} \right]$, $c_{i} \left(i=
 \overline{1,rN} \right)$ are the parameters representing the
 zonotope, and the matrix pair $ \{A,B \}$ is called the generator pair of the zonotopes.
 \end{defi}
 
 According to the above definition, the controllable region of the LDT systems is a special zonotope generated by the vectors of the matrix $P_{N} ^d$.

 As discusing above, simlilar to the definition of the braoad controllable region $R_N^d(A)$,
we can define the $N$-steps narrow controllable region $R^c_N(A)$ as follows
 \begin{defi} \label{de:d22020103} The $N$-steps narrow controllable region generated by the matrix pair $ \{A,B \}$ in the LDT models and the parameter set with a finite interval are defined as
 \begin{align}
 R_N^c(A) & = \left \{ \sum_{k=0} ^{N-1} A^{-N+k} B u_k, \quad \left \Vert u_k \right \Vert _ \infty \le1 \right \} \label{eq:e220206}
 \end{align}
 \end{defi}

 As pointed in paper \cite {zhaomw202004}, the $N$-steps narrow controllable region $R^c_N(A)$ and the reachable region $R^d_N(A)$ for the LDT systems $ \Sigma(A,B)$ satisfy the following relations
 \begin{align}
 R^c_N(A)& =A^{-1}R^d _N \left(A^{-1} \right) \label{eq:e220207} \\
 R^d_N(A)& =A^{-1}R^c _N \left(A^{-1} \right) \label{eq:e220208}
 \end{align}
 And then, their volumes satisfy the following relations
 \begin{align}
 V_n \left ( R^c_N(A) \right) & = \left \vert \det A \right \vert ^{-1} V_n \left (R^d _N \left(A^{-1} \right) \right) \label{eq:e220209} \\
 V_n \left (R^d_N(A) \right)& = \left \vert \det A \right \vert ^{-1} V_n \left ( R^c _N \left(A^{-1} \right) \right) \label{eq:e220210}
 \end{align}
where $V_n$ means the volume of the $n$-D regions. Therefore, the analysis and computing results of the regions $R^c_N(A)$ and $R^d_N(A)$ can be generalized conveniently to each others.

 \subsection {The recurssive volume computation of the finite-time controllable region } 

 By the linear transformation $ \overline x =Wx$ between two state vectors $x$ and $ \overline x$, the system models $ \Sigma(A,B)$ can be transformated as 
 $$ \Sigma \left( \overline A, \overline B \right)  = \Sigma \left( WAW^{-1},WB\right) $$
 \noindent and then the volumes of two controllable regions $R_N(A)$ and $R_N \left( \overline A \right)$ satisfy the following equation \cite{zhaomw202001}
 \begin{equation}
 V_n(R_N \left( \overline A) \right) = \left| \det W \right| V_n \left (R_N(A) \right) \label{eq:e220211}
 \end{equation}

When there exists some multiple eigenvalues or not in the system matrix $A$, the matrix $A$ can be transformated as Jordan or diagonal matrix by the linear transformation. And then, based on two kinds of special structure matrices $A$ and the transformating equation \eqref{eq:e220211}, some more effective methods for the volume computation of the controllable regions can be got.

When all eigenvalue of the matrix $A$ for the singel-input LDT systems are real numbers and different from each other, the diagonal tranformated systems can be represented as follows \cite{Kailath1980}, \cite{Chen1998}
 \begin{align}
 \Lambda & =W_dAW_d^{-1}= \textrm{diag} \left \{ \lambda_{1}, \lambda_{2}, \cdots, \lambda_{n} \right \} \label{eq:e220212} \\
 \Gamma & =W_dB= \left[ \beta_{1}, \beta_{2}, \cdots, \beta_{n} \right]^{T} \label{eq:e220213}
 \end{align}
where $ \beta_{i}=q_ib$ and $q_i$ is the $i$-th row of the diagonal transformating matrix $W_d$ and can be choosed as the unit left eigenvector corresponding to the $i$-th eigenvalue $ \lambda_{i}$.

It can be proven that the volumes of the zonotopes $R_N(A)$ and $R_N \left( \Lambda \right)$ can be computed recursively with complexity $ \mathcal{O}(N)$, and the corresponding result can be determined by the following theorem \cite{zhaomw202001}.
 
 \begin{theorem} \label {th:t22020102} If
 $ \Lambda$ is a diagonal matrix that all diagonal elements are differential each other and are with same signs, and $ \Gamma$ is only a vector, the volume of the zonotope $R_N \left( \Lambda \right)$ generated by matrix pair $ \{ \Lambda, \Gamma \}$ can be computed with computational complexity $ \mathcal{O}(N)$ by the following equation:
 \begin{equation}
 V_n \left(R_N \left( \Lambda \right) \right) = 2^n \left \vert \prod_{i=1}^{n} \beta_{i} \right \vert \left \vert V_{N}^{ \lambda_{1} \lambda_{2} \cdots \lambda_{n}} \right \vert \label{eq:e220214}
 \end{equation}
 where
 \begin{align}
 & V_{N}^{ \lambda_{1} \lambda_{2} \cdots \lambda_{n}}
 = \sum_{(i_{1},i_{2}, \cdots,i_{n}) \in \Omega_{0,N-1}^{n}}F_{ \lambda_{1} \lambda_{2} \cdots \lambda_{n}}^{i_{1}i_{2} \cdots i_{n}} \notag \\
 & \qquad \qquad =V_{N-1}^{ \lambda_{1} \lambda_{2} \cdots \lambda_{n}}+ \sum_{j=1}^{n}(-1)^{n+j} \lambda_{j}^{N-1}V_{N-1}^{ \lambda_{1} \lambda_{2} \cdots \lambda_{n} \setminus \lambda_{j}}, \quad N>n \label{eq:e220215} \\
 & V_{N}^{ \lambda_{i} } = \frac {1- \lambda_{i} ^N } {1- \lambda_{i} }, \qquad i=1,2, \dots,n \label{eq:e220216} \\
 & V_{n}^{ \lambda_{1} \lambda_{2} \cdots \lambda_{n}} = 
 F_{ \lambda_{1} \lambda_{2} \cdots \lambda_{n}}^{0,1, \cdots,n-1} = \prod_{1 \leq i<j \leq n} \left ( \lambda_{j}- \lambda_{i} \right )
 \label{eq:e220217} \\
 & F_{ \lambda_{1} \lambda_{2} \cdots \lambda_{n}}^{i_1 i_{2} \cdots i_{n}} = \det \left[ \begin{array}{cccc}
 \lambda_{1}^{i_{1}} & \lambda_{1}^{i_{2}} & \cdots & \lambda_{1}^{i_{n}} \\
 \lambda_{2}^{i_{1}} & \lambda_{2}^{i_{2}} & \cdots & \lambda_{2}^{i_{n}} \\
 \vdots & \vdots & \ddots & \vdots \\
 \lambda_{n}^{i_{1}} & \lambda_{n}^{i_{2}} & \cdots & \lambda_{n}^{i_{n}}
 \end{array} \right] \label{eq:e220218}
 \end{align}
 where '$ \lambda_{1} \lambda_{2} \cdots \lambda_{n} \setminus \lambda_{j}$' means that $ \lambda_{j}$ is deleted from sequence '$ \lambda_{1} \lambda_{2} \cdots \lambda_{n}$'.
 \end{theorem}

 \subsection {The analytical volume computation of the infinite-time controllable region for the matrix $A$ with $n$ differtential eigenvalues}

 For the needs of many analysis problems on the control capability of practical dynamic systems \eqref{eq:e220204}, our focus will be on the infinite-time controllable region $R_{ \infty} ^c (A)$ and reachable region $R_{ \infty} ^d (A) $. When the time variable $N \rightarrow \infty$, the computational cost of these region volumes by \textbf{Theorem \ref{th:t22020101}} and \textbf{Theorem \ref{th:t22020102}} will approach infinity and will exceed the accepted time cost in the analysis, design and on-line control for the practical engineering systems. We now propose a theorem on an analytic computation method with complexity $ \mathcal{O}(1)$ that has nothing to do with the time variable $N$ as follows \cite{zhaomw202001}.

 \begin{theorem} \label{th:t22020103}
 When the $n$ eigenvalues $ \lambda_i(i= \overline{1,n})$ of the matrix $A$ are different each other and are with same signs, the volume of the infinite-time $R_ \infty (A)$ is as
 \begin{equation} \label{eq:e220219}
 V_n \left (R_ \infty(A) \right) =2^n \left| \det W_d \right| ^{-1} \left \vert \prod_{i=1}^{n} \beta_{i} \right \vert \left| \Phi_{ \lambda_{1} \lambda_{2} \cdots \lambda_{n}} \right|
 \end{equation}
 where
 \begin{equation} \label{eq:e220220}
 \Phi_{ \lambda_{1} \lambda_{2} \cdots \lambda_{n}}= \left( \prod_{1 \leq j_{1}<j_{2} \leq n} \frac{ \lambda_{j_{2}}- \lambda_{j_{1}}}{1- \lambda_{j_{1}} \lambda_{j_{2}}} \right) \left( \prod_{i=1}^{n} \frac{1}{1- \left \vert \lambda_{i} \right \vert } \right)
 \end{equation}
 \end{theorem}

 \subsection{Decoding the Controllable Region}

According to the volume computing equation \eqref{eq:e220219}, some factors described the shape and size of the controllable region $R_N^d(A)$, that is, the control capability of the dynamical systems, are deconstructed as follows.
 \begin{align} 
 F_1 & = \left \vert \prod_{1 \leq j_{1}<j_{2} \leq n} \frac{ \lambda_{j_{2}}- \lambda_{j_{1}} }{ 1- \lambda_{j_{1}} \lambda_{j_{2}}} \right \vert \label{eq:e220221} \\
 F_1 ^{i,j}& = \left \vert \frac{ \lambda_{j}- \lambda_{i} }{ 1- \lambda_{j} \lambda_{i}} \right \vert \label{eq:e220222} \\
 F _{ \left( j_1 \dots j_s \right)} ^{ \left( k_1 \dots k_q \right) } & = \prod_{j_i \in \left(j_{1}, \dots,j_{s} \right)}
 \prod_{k_v \in \left(k_{1} \dots k_{q} \right)} \frac{1- \lambda_{j_{i}} \lambda_{k_{v}}}{ \lambda_{k_{v}}- \lambda_{j_{i}}} \label {eq:e220223} \\
 F_{2,i} 
 &= \frac{ \left \vert q_ib \right \vert}{ 1- \left \vert \lambda_{i} \right \vert }, \; \;i=1,2, \dots,n
 \label{eq:e220224} \\
 F_{3,i} & = \left \vert q_ib \right \vert, \; \;i=1,2, \dots,n \label{eq:e220225} 
 \end{align}
The above analytical factors can be called respectively as 
the shape(pole distribution factor) factor, the shape factor in the 2-D setcion of the region, the side length of the circumscribed rhombohedral, and the modal controllability. In fact, the shape factor $F_1$ is also the eigenvalue evenness factor of the linear system, and can describe the control capability caused by the eigenvalue distribution. In addition, the modal controllability factor $F_{3,i}$ have been put forth by papers \cite{chan1984} \cite{HamElad1988} \cite{ChoParLee2000} \cite{Chenliu2001}, and will not be discussed here.

 \section {Some Key Lemmas }

 First, for that matrix pair, the following lemma about the sign of a class of quasi-Vandermonde matrices is proposed and proven.

 \begin{lemma} \label{le:L22020101} For any $n>0$, if $0 \le k_{1}<k_{2}< \cdots<k_{n}$ and $0< \lambda_{1}< \lambda_{2}< \cdots< \lambda_{n}$, we have
 \begin{equation} \label{eq:e220226}
 F_{ \lambda_{1} \lambda_{2} \cdots \lambda_{n}}^{k_1,k_{2} \cdots k_{n}} = \det \left[ \begin{array}{cccc}
 \lambda_{1}^{k_{1}} & \lambda_{1}^{k_{2}} & \cdots & \lambda_{1}^{k_{n}} \\
 \lambda_{2}^{k_{1}} & \lambda_{2}^{k_{2}} & \cdots & \lambda_{2}^{k_{n}} \\
 \vdots & \vdots & \ddots & \vdots \\
 \lambda_{n}^{k_{1}} & \lambda_{n}^{k_{2}} & \cdots & \lambda_{n}^{k_{n}}
 \end{array} \right] >0
 \end{equation}
 \end{lemma}
 
 \begin{lemma} \label{le:L22020102} For any variables $ \lambda _i \in C \left( i= \overline {1,n} \right)$, the following equation holds.
 \begin{equation} \label{eq:e220227}
 \left(1- \Upsilon_n \right) \Phi_{ \lambda_{1} \cdots \lambda_{n}} = \sum_{k=1}^{n}(-1)^{1+k} \Upsilon_{n \setminus k} \Phi_{ \lambda_{1} \cdots \lambda_{n} \setminus \lambda_{k}}
 \end{equation}
 \end{lemma}
 where '$ \lambda_{1} \cdots \lambda_{n} \setminus \lambda_{j}$' means that $ \lambda_{j}$ is deleted from sequence $ \lambda_{1} \cdots \lambda_{n}$.
 \begin{align}
& \Upsilon_s = \prod_{i=1}^{s} \lambda_{i} , \quad \Upsilon_{s \setminus k} = \prod_{i={ \overline {1,s} \setminus k}} \lambda_{i} \label{eq:e220228} \\
& \Phi_{ \lambda_{1} \cdots \lambda_{n}}= \left( \prod_{1 \leq j<k \leq n} \frac{ \lambda_{k}- \lambda_{j}}{1- \lambda_{j} \lambda_{k}} \right) \left( \prod_{i=1}^{n} \frac{1}{1- \lambda_{i}} \right) \label{eq:e220229}
 \end{align}
 
 \begin{lemma} \label{le:L22020103} For the any set $ \Lambda = \left \{ \lambda_{k_{1}}, \lambda_{k_{2}}, \dots, \lambda_{k_{m}} \right \}$ and the distribution factor $ \Phi_ \Lambda$, the following three equations hold.
 \begin{align}
 &(1) \; \left. \Phi_{ \Lambda} \right|_{ \lambda_{j}= \lambda_{i}}= \left \{ \begin{array}{ll}
 0 & \lambda_i, \lambda_j \in \Lambda \\
 (-1)^{s-h-1} \Phi_{ \lambda_{i} \cup \Lambda \backslash \lambda_{j}} & k_h< i < k_{h+1} \le j=k_{s} \\
 \Phi_{ \Lambda} & \lambda_j \notin \Lambda
 \end{array} \right. \label{eq:e220230}
 \\
&(2) \; \left. \left(1- \lambda_{i} \right) \Phi_{ \Lambda} \right|_{ \lambda_{i}=1}= \left \{ \begin{array}{ll}
 (-1)^{m-h} \Phi_{ \Lambda \backslash \lambda_{i}} & i=k_{h} \\
 0 & \textrm{others}
 \end{array} \right. \label{eq:e220231}
 \\
&(3) \; \left. \left(1- \lambda_{i} \lambda_{j} \right) \Phi_{ \Lambda} \right|_{ \lambda_{j}= \frac{1}{ \lambda_{i}}}= \left \{ \begin{array}{ll}
 (-1)^{s-h} \frac{1+ \lambda_{i}}{1- \lambda_{i}} \Phi_{ \Lambda \backslash \lambda_{i} \backslash \lambda_{j}} & i=k_{h}<j=k_{s} \\
 0 & \textrm{others}
 \end{array} \right. \label{eq:e220232}
 \end{align} 
 \end{lemma}
 
 \textbf{Lemmas \ref {le:L22020101}} and \textbf{ \ref {le:L22020102}} are got in papaer \cite{zhaomw202001}, \textbf{Lemma \ref {le:L22020103}} can be proven as follows.
 
 \textbf{Proof of Lemma \ref {le:L22020103}}. (1) Eq. \eqref {eq:e220230} can be proven by the definition equation \eqref{eq:e220229} as follows.

1) By Eq. \eqref {eq:e220229}, if $ \lambda_i$ and $ \lambda_j$ are in the sequence $K$, we have
 \begin{align}
 \left. \Phi_{ \Lambda} \right|_{ \lambda_{j}= \lambda_{i}}=0 \label{eq:e220233}
 \end{align}

2) By Eq. \eqref {eq:e220229}, if $k_h< i < k_{h+1} \le j=k_{s}$, we have
 \begin{align}
 \Phi_{ \Lambda} 
 & = \left( \prod_{1 \leq q<r \leq m} \frac{ \lambda_{k_r}- \lambda_{k_q}}{1- \lambda_{k_q} \lambda_{k_r}} \right) \left( \prod_{q=1}^{m} \frac{1}{1- \lambda_{k_q}} \right) \notag \\
 & = \frac { \Phi_{ \Lambda \backslash \lambda_{j}}}
 {1- \lambda_j }
 \left( \prod_{q=1 } ^{s-1} \frac{ \lambda_{j}- \lambda_{k_q}}{1- \lambda_{j} \lambda_{k_q}} \right)
 \left( \prod_{q=s+1} ^{m} \frac{ \lambda_{k_q}- \lambda_{j}}{1- \lambda_{k_q} \lambda_{j}} \right) \label{eq:e220234}
 \end{align}
And then, we have
 \begin{align}
 \left. \Phi_{ \Lambda} \right|_{ \lambda_{j}= \lambda_{i}} 
 & = \frac { \Phi_{ \Lambda \backslash \lambda_{j}}}
 {1- \lambda_i }
 \left( \prod_{q=1 } ^{s-1} \frac{ \lambda_{i}- \lambda_{k_q}}{1- \lambda_{i} \lambda_{k_q}} \right)
 \left( \prod_{q=s+1} ^{m} \frac{ \lambda_{k_q}- \lambda_{i}}{1- \lambda_{k_q} \lambda_{i}} \right) \notag 
 \\
 & =(-1)^{s-h-1} \frac { \Phi_{ \Lambda \backslash \lambda_{j}}}
 {1- \lambda_i }
 \left( \prod_{q=1 } ^{h} \frac{ \lambda_{i}- \lambda_{k_q}}{1- \lambda_{i} \lambda_{k_q}} \right)
 \left( \prod_{q=h+1} ^{m \backslash s} \frac{ \lambda_{k_q}- \lambda_{i}}{1- \lambda_{k_q} \lambda_{i}} \right) \notag 
 \\
 &=(-1)^{s-h-1} \Phi_{ \lambda_{i} \cup \Lambda \backslash \lambda_{j}} \label{eq:e220235}
 \end{align}

3) if $ \lambda_j$ isn't in the sequence $K$, we have 
 \begin{align}
 \left. \Phi_{ \Lambda} \right|_{ \lambda_{j}= \lambda_{i}}= \Phi_{ \Lambda} \label{eq:e220236}
 \end{align}

(2) By Eq. \eqref {eq:e220229} and \eqref {eq:e220234}, if $i=k_h$, we have
 \begin{align}
 \left. \left(1- \lambda_{i} \right) \Phi_{ \Lambda} \right|_{ \lambda_{i}=1} 
& = \Phi_{ \Lambda \backslash \lambda_{i}} \left.
 \left ( \prod_{q=1} ^{h-1} \frac{ \lambda_{i}- \lambda_{k_q}}{1- \lambda_{k_q} \lambda_{i}} \right )
 \left ( \prod_{q=h+1} ^{m} \frac{ \lambda_{k_q}- \lambda_{i} }{1- \lambda_{k_q} \lambda_{i}} \right )
 \right \vert _{ \lambda_{i}=1} \notag \\
& = (-1)^{s-h} \Phi_{ \Lambda \backslash \lambda_{i}} \label{eq:e220237}
 \end{align}

If $ \lambda_i \notin \Lambda$, we have,
 \begin{align}
 \left. \left(1- \lambda_{i} \right) \Phi_{ \Lambda} \right|_{ \lambda_{i}=1}=0 \label{eq:e220238}
 \end{align}

(3) By Eq. \eqref {eq:e220229}, if $ i=k_{h}<j=k_{s}$, we have
 \begin{align}
 \Phi_{ \Lambda} & = \frac { \left ( \lambda_j- \lambda_i \right) \Phi_{ \Lambda \backslash \lambda_{i} \backslash \lambda_{j} }}{ \left (1- \lambda_i \right) \left (1- \lambda_j \right) \left (1- \lambda_i \lambda_j \right)} \left ( \prod_{q=1} ^{h-1} \frac{ \lambda_{i}- \lambda_{k_q}}{1- \lambda_{k_q} \lambda_{i}} \right )
 \left ( \prod_{q=h+1} ^{m \backslash s} \frac{ \lambda_{k_q}- \lambda_{i} }{1- \lambda_{k_q} \lambda_{i}} \right ) \notag \\
& \quad \times \left ( \prod_{q=1} ^{s-1 \backslash h} \frac{ \lambda_{j}- \lambda_{k_q}}{1- \lambda_{k_q} \lambda_{j}} \right )
 \left ( \prod_{q=s+1} ^{m} \frac{ \lambda_{k_q}- \lambda_{j} }{1- \lambda_{k_q} \lambda_{j}} \right ) \label{eq:e220239}
 \end{align}
Therefore, we have
 \begin{align}
 \left. \left(1- \lambda_{i} \lambda_{j} \right) \Phi_{ \Lambda} \right \vert _{ \lambda_{j}= \frac {1} { \lambda_{i}}} 
 & = \frac { \left (1- \lambda_i ^2 \right) \Phi_{ \Lambda \backslash \lambda_{i} \backslash \lambda_{j} }}{- \left (1- \lambda_i \right) ^2 } \left ( \prod_{q=1} ^{h-1} \frac{ \lambda_{i}- \lambda_{k_q}}{1- \lambda_{k_q} \lambda_{i}} \right )
 \left ( \prod_{q=h+1} ^{m \backslash s} \frac{ \lambda_{k_q}- \lambda_{i} }{1- \lambda_{k_q} \lambda_{i}} \right ) \notag \\
 & \quad \times \left ( \prod_{q=1} ^{s-1 \backslash h} \frac{ 1- \lambda_{i} \lambda_{k_q}}{ \lambda_{i} - \lambda_{k_q} } \right )
 \left ( \prod_{q=s+1} ^{m} \frac{ \lambda_{i} \lambda_{k_q}-1 }{ \lambda_{i} - \lambda_{k_q} } \right) \notag \notag \\
 & = (-1)^{s-h} \frac{1+ \lambda_{i}}{1- \lambda_{i}} \Phi_{ \Lambda \backslash \lambda_{i} \backslash \lambda_{j}} \label{eq:e220240}
 \end{align}

In addition, if $ \lambda _i \notin \Lambda$ or $ \lambda _j \notin \Lambda$, we have 
 \begin{align}
 \left. \left(1- \lambda_{i} \lambda_{j} \right) \Phi_{ \Lambda} \right \vert _{ \lambda_{j}= \frac {1} { \lambda_{i}}} =0 \label{eq:e220241}
 \end{align}
 \qed

 \section {The analytical volume computation of the finite-time controllable region}
 Based on the above lemma, we have the following analytical theorem of the volume computation about the finite-time controllable region $R_N(A)$.
 
 \begin{theorem} \label{th:t22020104}
 If the $n$ eigenvalues $ \lambda_i(i= \overline{1,n})$ of the matrix $A$ are positive real numbers and satisfy 
 $$
 0< \lambda_1< \lambda_2< \cdots< \lambda_n,
 $$
the volume of the finite-time controllable region $R_N(A)$ when $N \ge n$ can be computed analytical as follows
 \begin{align}
 V_n \left(R_N(A) \right) &=2^n \left \vert \det \left ( W_d^{-1} \right) \prod _{i=1} ^n \beta_i \right \vert V_{N}^{ \lambda_{1} \cdots \lambda_{n}} \label{eq:e220242} \\
V_{N}^{ \lambda_{1} \cdots \lambda_{n}}
& = \sum_{ \left(i_{1}, \dots,i_{n} \right) \in \Omega_{0,N-1} ^n}
 F_{ \lambda_{1} \cdots \lambda_{n}} ^{i_{1} \dots i_{n}} \notag
 \\
& = \Phi_{ \lambda_{1} \cdots \lambda_{n}} + \sum_{s=1} ^n \sum _{ \left( j_1, \dots,j_s \right) \in \varTheta_n ^s} 
c^{(n)} _{ \left( j_1 \dots j_s \right) } \Upsilon _{ \left( j_1 \dots j_s \right)} ^N \Phi _{ \left( j_1 \dots j_s \right) } 
 \Phi _{ n \setminus \left( j_1 \dots j_s \right) } \notag
 \\
 & = \sum_{s=0} ^n \sum _{ \left( j_1, \dots,j_s \right) \in \varTheta_n ^s} 
 c^{(n)} _{ \left( j_1 \dots j_s \right) } \Upsilon _{ \left( j_1 \dots j_s \right)} ^N \Phi _{ \left( j_1 \dots j_s \right) } 
 \Phi _{ n \setminus \left( j_1 \dots j_s \right) } \label{eq:e220243}
 \\
& = \sum_{s=0} ^n \sum _{ \left( j_1, \dots,j_s \right) \in \varTheta_n ^s} 
c^{(n)} _{ n \setminus \left( j_1 \dots j_s \right) } \Upsilon _{ n \setminus \left( j_1 \dots j_s \right)} ^N \Phi _{ \left( j_1 \dots j_s \right) } 
 \Phi _{ n \setminus \left( j_1 \dots j_s \right) } \label{eq:e220244}
 \\
&= \left \{ \sum_{s=0}^{n} \sum_{ \left(j_{1}, \dots,j_{s} \right) \in \varTheta_{n}^{s}}c_{ \left(j_{1} \dots j_{s} \right)}^{(n)} \Upsilon_{ \left(j_{1} \dots j_{s} \right)}^{N} F_{ \left(j_{1} \dots j_{s} \right)} ^{n \setminus \left(j_{1} \dots j_{s} \right)} \right \} 
 \Phi_{ \left(1 \dots n \right)}  \label{eq:e220245} 
 \end{align}
where $ F_{ \left(j_{1} \dots j_{s} \right)} ^{n \setminus \left(j_{1} \dots j_{s} \right)}$ is defined as Eq. \eqref{eq:e220223};
 the $s$-tuple set $ \varTheta_{n}^{s}$ consists of all possible $s$-tuples
 $(j_{1}, \dots,j_{s})$
 whose elements are picked from the set
 $ \left \{ 1,2, \cdots,n \right \}$
 and sorted by their values;
 \begin{align}
 & c^{(n)} _{ \left( j_1 \dots j_s \right) }=(-1)^{(n+1)s- \sum_{i=1}^{s}j_{i}} \label{eq:e220246} \\
 & \Upsilon _{ \left( j_1 \dots j_s \right)} ^N = \prod_{k=1} ^{ s} \lambda _{j_k} ^N \label{eq:e220247} \\
 & \Phi _{ \left( j_1 \dots j_s \right) } = \left( \prod_{1 \leq i<k \leq s} \frac{ \lambda_{j_{k}}- \lambda_{j_{i}}}{1- \lambda_{j_{i}} \lambda_{j_{k}}} \right) \left( \prod_{i=1}^{s} \frac{1}{1- \lambda_{j_i}} \right) \label{eq:e220248}
 \end{align} 
where $c^{(n)} _{- }= 1$, $ \Upsilon _{ -} ^N=1$, and $ \Phi _{ - }=1$; "-" means that the number sequence is blank. 
 \end{theorem}

By the above theorem, if all eighenvalues $ \lambda_i$ are in $[0,1)$, the function $V_{N}^{ \lambda_{1} \cdots \lambda_{n}}$ can be represented as follows
 \begin{align}
 V_{N}^{ \lambda_{1} \cdots \lambda_{n}}= \Phi_{ \lambda_{1} \cdots \lambda_{n}} + \sum _{i=0} ^n O( \lambda _i^N) \label{eq:e220249}
 \end{align}
where $ \lim _{N \rightarrow \infty } O( \lambda _i^N)=0$, that is, $V_{ \infty}^{ \lambda_{1} \cdots \lambda_{n}}= \Phi_{ \lambda_{1} \cdots \lambda_{n}} $. Therefore, the anyltical volume equation of the infinite-time controllable region $R_N^d(A)$ in papers \cite {zhaomw202001} and \cite {zhaomw202004} is a special example of the above theorem.

 \textbf{Proof of Theorem \ref {th:t22020104}}. The theorem can be proven by the inductive method. Firstly, Eq. \eqref {eq:e220243} when $N=n$ is proven by the first-time inductive method. And then, based on the proven result when $N=n$, Eq. \eqref {eq:e220243} when $N>n$ is proven by the second-time inductive method.
 
 (1) Next, Eq. \eqref {eq:e220243} when $N=n$ is proven at first.
 
1.1) When $n=2$, by the definition of $V_{N}^{ \lambda_{1} \lambda_{2}}$ and $F_{ \lambda_{1} \lambda_{2}} ^{i_{1} i_{2} }$, we have 
 \begin{align} 
 V_{N}^{ \lambda_{1} \lambda_{2}}
 & = \sum_{ \left(i,j \right) \in \Omega_{0,N-1} ^2}
 F_{ \lambda_{1} \lambda_{2}} ^{i j } = \sum _{i=0} ^{N-2} \sum _{j=i+1}^{N-1} \left| \begin{array}{cc} 
 \lambda_{1}^{i} & \lambda_{1}^{j} \\
 \lambda_{2}^{i} & \lambda_{2}^{j}
 \end{array} \right| \notag
 \\ 
 & = \sum _{i=0} ^{N-2} \sum _{j=i+1}^{N-1} \left [ \lambda_{1}^{i} \lambda_{2}^{j} - \lambda_{1}^{j} \lambda_{2}^{i} \right] \notag
 \\ 
 & = \sum _{i=0} ^{N-2} \left [ \lambda_{1}^{i} \frac { \lambda_{2}^{i+1}- \lambda_{2}^{N} } {1- \lambda_{2}} - \lambda_{2}^{i} \frac { \lambda_{1}^{i+1}- \lambda_{1}^{N} } {1- \lambda_{1}} \right] \notag
 \\ 
 & = \sum _{i=0} ^{N-2} \frac { \lambda_{1}^{i} \left( \lambda_{2}^{i+1}- \lambda_{2}^{N} \right) \left(1- \lambda_{1} \right) 
 - \lambda_{2}^{i} \left( \lambda_{1}^{i+1}- \lambda_{1}^{N} \right) \left(1- \lambda_{2} \right) } { \left(1- \lambda_{1} \right) \left(1- \lambda_{2} \right)} \notag
 \\ 
& = \sum _{i=0} ^{N-2} \frac { \left[ \lambda_{2} \left(1- \lambda_{1} \right) 
 - \lambda_{1} \left(1- \lambda_{2} \right) \right ] \lambda_{1}^{i} \lambda_{2}^{i} 
 - \lambda_{1}^{i} \lambda_{2}^{N} \left(1- \lambda_{1} \right) 
 + \lambda_{2}^{i} \lambda_{1}^{N} \left(1- \lambda_{2} \right) } { \left(1- \lambda_{1} \right) \left(1- \lambda_{2} \right)} \notag
 \\ 
 & = \frac { \left( \lambda_{2}- \lambda_{1} \right) \frac {1- \lambda_{1}^{N-1} \lambda_{2}^{N-1} } {1- \lambda_{1} \lambda_{2} }- \left ( 1- \lambda_{1}^{N-1} \right) \lambda_{2}^{N} 
 + \left ( 1- \lambda_{2}^{N-1} \right) \lambda_{1}^{N} } { \left(1- \lambda_{1} \right) \left(1- \lambda_{2} \right)} \notag
 \\ 
& = \frac { \left( \lambda_{2}- \lambda_{1} \right) \left ( 1- \lambda_{1}^{N-1} \lambda_{2}^{N-1} \right)
 } { \left(1- \lambda_{1} \right) \left(1- \lambda_{2} \right) \left( 1- \lambda_{1} \lambda_{2} \right) }
 - \frac { \lambda_{2}^{N} - \lambda_{1}^{N} - \left ( \lambda_{2}- \lambda_{1} \right) \lambda_{1}^{N-1} \lambda_{2}^{N-1} 
 } { \left(1- \lambda_{1} \right) \left(1- \lambda_{2} \right)} \notag
 \\ 
& = \frac { \left( \lambda_{2}- \lambda_{1} \right) \left ( 1- \lambda_{1}^{N} \lambda_{2}^{N} \right)
} { \left(1- \lambda_{1} \right) \left(1- \lambda_{2} \right) \left( 1- \lambda_{1} \lambda_{2} \right) }
- \frac { \lambda_{2}^{N} - \lambda_{1}^{N} } { \left(1- \lambda_{1} \right) \left(1- \lambda_{2} \right)} \label{eq:e220250}
 \end{align} 
 
 By Eqs. \eqref {eq:e220246}, \eqref {eq:e220247} and \eqref{eq:e220248}, when $n=2$, we have
 \begin{align} 
 \sum_{s=0} ^2 & \sum _{ \left( j_1, \dots,j_s \right) \in \varTheta_2 ^s} 
c^{(2)} _{ \left( j_1 \dots j_s \right) } \Upsilon _{ \left( j_1 \dots j_s \right)} ^N \Phi _{ \left( j_1 \dots j_s \right) } 
 \Phi _{ 2 \setminus \left( j_1 \dots j_s \right) } \notag
 \\
& =c^{(2)} _{ - } \Phi _{-} \Phi_{ \lambda_{1} \lambda_{2} } 
 + c^{(2)} _{ 1 } \Upsilon _{ 1 } ^N \Phi _{ 1 } 
 \Phi _{2 } 
 + c^{(2)} _{ 2 } \Upsilon _{ 2 } ^N \Phi _{ 2 } 
 \Phi _{1 } 
 +c^{(2)} _{ 12 } \Upsilon _{ 12 } ^N \Phi _{ 12 } 
 \Phi _{-} \notag
 \\
& = \frac { \left( \lambda_{2}- \lambda_{1} \right) \left ( 1- \lambda_{1}^{N} \lambda_{2}^{N} \right)
} { \left(1- \lambda_{1} \right) \left(1- \lambda_{2} \right) \left( 1- \lambda_{1} \lambda_{2} \right) }
- \frac { \lambda_{2}^{N} - \lambda_{1}^{N} } { \left(1- \lambda_{1} \right) \left(1- \lambda_{2} \right)} \label{eq:e220251}
 \end{align}

Therefore, by Eqs. \eqref {eq:e220250} and \eqref{eq:e220251}, we can see, Eq. \eqref {eq:e220243} holds when $n=2$.
 \\
It is worth noting that, when $N=n=2$, we have
 \begin{align} V_{2}^{ \lambda_{1} \lambda_{2}} 
 & = \frac { \left( \lambda_{2}- \lambda_{1} \right) \left ( 1- \lambda_{1}^{2} \lambda_{2}^{2} \right)
 } { \left(1- \lambda_{1} \right) \left(1- \lambda_{2} \right) \left( 1- \lambda_{1} \lambda_{2} \right) }
 - \frac { \lambda_{2}^{2} - \lambda_{1}^{2} } { \left(1- \lambda_{1} \right) \left(1- \lambda_{2} \right)} \notag \\
 & = \lambda_{2} - \lambda_{1} \label{eq:e220252}
 \end{align}

(2) Considered that
 \begin{align} 
 V_{n}^{ \lambda_{1} \cdots \lambda_{n}}
 = F_{ \lambda_{1} \cdots \lambda_{n}}^{01 \cdots n}
 = \left| \begin{array}{cccc} 
 1 & \lambda_{1} & \dots & \lambda_{1}^{n} \\
 1 & \lambda_{2} & \dots & \lambda_{2}^{n} \\
 \vdots & \vdots & \ddots & \vdots \\
 1 & \lambda_{n} & \dots & \lambda_{n}^{n}
 \end{array} \right |
 = \prod_{1 \le i < j \le n} \left( \lambda_{j} - \lambda_{i} \right)
 \label{eq:e220253} 
 \end{align}
to prove that Eq. \eqref {eq:e220243} holds when $N=n$, it is needs to prove that for the rational function
 \begin{align}
G_{ \lambda_{1} \cdots \lambda_{n}} = \sum_{s=0} ^n \sum _{ \left( j_1, \dots,j_s \right) \in \varTheta_n ^s} 
 c^{(n)} _{ \left( j_1 \dots j_s \right) } \Upsilon _{ \left( j_1 \dots j_s \right)} ^n \Phi _{ \left( j_1 \dots j_s \right) } 
 \Phi _{ n \setminus \left( j_1 \dots j_s \right) } 
 \label{eq:e220254}
 \end{align}
satisfies the following equation 
 \begin{align}
G_{ \lambda_{1} \cdots \lambda_{n}}
 = \prod_{1 \le i < j \le n} \left( \lambda_{j} - \lambda_{i} \right)
 \label{eq:e220255}
 \end{align}
By the definition of the functions $\Phi$  and Eq. \eqref{eq:e220254}, we know, the factors $1- \lambda_i$ and $1- \lambda_j \lambda_k$ for $1 \le i \le n$ and $1 \le j < k \le n$ are the all factors in the denominator polynomial of the rational function $G_{ \lambda_{1} \cdots \lambda_{n}}$. Therefore, Eq. \eqref{eq:e220255} is equal to the following equation
 \begin{align}
& G_{ \lambda_{1} \cdots \lambda_{n}} \prod _{i=0} ^{n} \left( 1- \lambda_i \right) \times \prod_{1 \le i < j \le n} \left( 1- \lambda_{i} \lambda_{j} \right)
 \notag \\
& \qquad = \prod _{i=0} ^{n} \left( 1- \lambda_i \right) \times \prod_{1 \le i < j \le n} \left [ \left( \lambda_{j} - \lambda_{i} \right) \left( 1- \lambda_{i} \lambda_{j} \right) \right]
 \label{eq:e220256}
 \end{align}
Hence, to prove that Eq. \eqref {eq:e220255}holds, it is needs only to prove the following equations hold for all possible $i$ and $j$.
 \begin{align}
 \left. G_{ \lambda_{1} \cdots \lambda_{n}} \right | _{ \lambda_j = \lambda_i} & =0 \label{eq:e220257} \\
 \left. \left( 1- \lambda_i \right) G_{ \lambda_{1} \cdots \lambda_{n}} \right | _{ \lambda_i =1} & =0  \label{eq:e220258}\\
 \left. \left( 1- \lambda_i \lambda_j \right) G_{ \lambda_{1} \cdots \lambda_{n}} \right | _{ \lambda_j = \frac {1} { \lambda_i}} & =0 
 \label{eq:e220259}
 \end{align}
In fact, the function $G_{ \lambda_{1} \cdots \lambda_{n}} $ canbe represented as follows
 \begin{align}
 G_{ \lambda_{1} \cdots \lambda_{n}} & = \sum_{s=0} ^{n-2} \sum _{ \left( j_1, \dots,j_s \right) \in \varTheta_{n \setminus i \setminus j} ^s} \left[ 
 c^{(n)} _{ \left(i, j_1 \dots j_s \right) } \Upsilon _{ \left(i, j_1 \dots j_s \right)} ^n \Phi _{ \left(i, j_1 \dots j_s \right) } 
 \Phi _{ n \setminus i \setminus \left( j_1 \dots j_s \right) } \right . \notag
 \\
 & \quad + c^{(n)} _{ \left(j, j_1 \dots j_s \right) } \Upsilon _{ \left(j, j_1 \dots j_s \right)} ^n \Phi _{ \left(j, j_1 \dots j_s \right) } 
 \Phi _{ n \setminus j \setminus \left( j_1 \dots j_s \right) } \notag
 \\
 & \quad + c^{(n)} _{ \left(i,j, j_1 \dots j_s \right) } \Upsilon _{ \left(i,j, j_1 \dots j_s \right)} ^n \Phi _{ \left(i,j, j_1 \dots j_s \right) } 
 \Phi _{ n \setminus i \setminus j \setminus \left( j_1 \dots j_s \right) } \notag
 \\
 & \quad \left. + c^{(n)} _{ \left(j_1 \dots j_s \right) } \Upsilon _{ \left( j_1 \dots j_s \right)} ^n \Phi _{ \left( j_1 \dots j_s \right) } 
 \Phi _{ n \setminus \left( j_1 \dots j_s \right) } \right ] \label {eq:e220260} \\
 \textnormal{or } & \notag \\
 & = \sum_{s=0} ^{n-1} \sum _{ \left( j_1, \dots,j_s \right) \in \varTheta_{n \setminus i } ^s} \left[ 
 c^{(n)} _{ \left(i, j_1 \dots j_s \right) } \Upsilon _{ \left(i, j_1 \dots j_s \right)} ^n \Phi _{ \left(i, j_1 \dots j_s \right) } 
 \Phi _{ n \setminus i \setminus \left( j_1 \dots j_s \right) } \right . \notag
 \\
 & \quad \left. + c^{(n)} _{ \left( j_1 \dots j_s \right) } \Upsilon _{ \left( j_1 \dots j_s \right)} ^n \Phi _{ \left( j_1 \dots j_s \right) } 
 \Phi _{ n \setminus \left( j_1 \dots j_s \right) } \label {eq:e220261}
 \right ]
 \end{align}

By \textbf{Lemma \ref{le:L22020103} } and Eq. \eqref {eq:e220260}, we have
 \begin{align}
 \left. G_{ \lambda_{1} \cdots \lambda_{n}} \right | _{ \lambda_j = \lambda_i} & = \sum_{s=0} ^{n-2} \sum _{ \left( j_1, \dots,j_s \right) \in \varTheta_{n \setminus i \setminus j} ^s} \left[ 
 c^{(n)} _{ \left(i, j_1 \dots j_s \right) } \Upsilon _{ \left(i, j_1 \dots j_s \right)} ^n \Phi _{ \left(i, j_1 \dots j_s \right) } 
 \Phi _{ n \setminus i \setminus \left( j_1 \dots j_s \right) } \right . \notag
 \\
 & \quad \left. + c^{(n)} _{ \left(j, j_1 \dots j_s \right) } \Upsilon _{ \left(j, j_1 \dots j_s \right)} ^n \Phi _{ \left(j, j_1 \dots j_s \right) } 
 \Phi _{ n \setminus j \setminus \left( j_1 \dots j_s \right) } \right ] _{ \lambda_j = \lambda_i} \notag
 \\ 
 & = \sum_{s=0} ^{n-2} \sum _{ \left( j_1, \dots,j_s \right) \in \varTheta_{n \setminus i \setminus j} ^s} \Upsilon _{ \left(i, j_1 \dots j_s \right)} ^n 
 \left[ (-1)^{(n+1)s-i- \sum_{k=1}^{s}j_{k}}
 \times (-1)^{h-1}
 \right . \notag
 \\
 & \quad \left. + (-1)^{(n+1)s-j- \sum_{k=1}^{s}j_{k}} \times (-1)^{j-i-h} 
 \right ] \Phi _{ \left(i, j_1 \dots j_s \right) }
 \Phi _{ n \setminus j \setminus \left( j_1 \dots j_s \right) } \notag
 \\ 
 & = \sum_{s=0} ^{n-2} \sum _{ \left( j_1, \dots,j_s \right) \in \varTheta_{n \setminus i \setminus j} ^s} \Upsilon _{ \left(i, j_1 \dots j_s \right)} ^n 
 (-1)^{(n+1)s- \sum_{k=1}^{s}j_{k}} \left[ (-1)^{-i+h-1}
 \right . \notag
 \\
 & \quad \left. + (-1)^{-i-h} 
 \right ] \Phi _{ \left(i, j_1 \dots j_s \right) }
 \Phi _{ n \setminus j \setminus \left( j_1 \dots j_s \right) } \notag
 \\
 &=0
 \label {eq:e220262}
 \end{align}
where $h$ is the position difference between the numbers $i$ and $j$ when $i$ and $j$ insert the ordered sequence $j_1j_2 \dots j_s$. Therefore, $j-i-h+1$ is also the position difference between $i$ and $j$ in the ordered sequence $(12 \dots n) \setminus j_1j_2 \dots j_s$. By the above deduction, we can see, for any $j \ne i$, $ \lambda_j - \lambda_i$ must be the factors of rational function $G_{ \lambda_{1} \cdots \lambda_{n}}$.

By \textbf{Lemma \ref{le:L22020103} } and Eq. \eqref {eq:e220261}, we have
 \begin{align}
 & \left. \left ( 1- \lambda_i \right) G_{ \lambda_{1} \cdots \lambda_{n}} \right | _{ \lambda_i=1} \notag 
 \\
 & \qquad = \sum_{s=0} ^{n-1} \sum _{ \left( j_1, \dots,j_s \right) \in \varTheta_{n \setminus i } ^s} \left[ 
 c^{(n)} _{ \left(i, j_1 \dots j_s \right) } \Upsilon _{ \left(i, j_1 \dots j_s \right)} ^n \left( 1- \lambda_i \right) \Phi _{ \left(i, j_1 \dots j_s \right) } 
 \Phi _{ n \setminus i \setminus \left( j_1 \dots j_s \right) } \right . \notag
 \\
 & \qquad \quad \left. + c^{(n)} _{ \left( j_1 \dots j_s \right) } \Upsilon _{ \left( j_1 \dots j_s \right)} ^n \Phi _{ \left( j_1 \dots j_s \right) } 
 \left( 1- \lambda_i \right) \Phi _{ n \setminus \left( j_1 \dots j_s \right) } \right] _{ \lambda_i=1} \notag 
 \\
 & \qquad = \sum_{s=0} ^{n-1} \sum _{ \left( j_1, \dots,j_s \right) \in \varTheta_{n \setminus i } ^s} \Upsilon _{ \left(j_1 \dots j_s \right)} ^n \left[ 
 (-1)^{(n+1)(s+1)-i- \sum_{k=1}^{s}j_{k}}
 \times \left( -1 \right) ^{s+1- \delta} 
 \right . \notag
 \\
 & \qquad \quad \left. + (-1)^{(n+1)s- \sum_{k=1}^{s}j_{k}}
 \times 
 \left(- 1 \right) ^ {n-s-i+ \delta-1} \right] \Phi _{ \left( j_1 \dots j_s \right) } \Phi _{ n \setminus i \setminus \left( j_1 \dots j_s \right) } \notag 
 \\
 & \qquad =0 \label {eq:e220263}
 \end{align}
where $ \delta$ is the position of the number $i$ when $i$ inserts the ordered sequence $j_1j_2 \dots j_s$. Therefore, $i- \delta+1$ is also the position of $i$ in the ordered sequence $(12 \dots n) \setminus j_1j_2 \dots j_s$. By the above deduction, we can see, for any $ i$, $1- \lambda_i$ must be the factors in the denominator polynomial of rational function $G_{ \lambda_{1} \cdots \lambda_{n}}$.

By \textbf{Lemma \ref{le:L22020103} } and Eq. \eqref {eq:e220260}, when $j>i$, we have,
 \begin{align}
& \left. \left ( 1- \lambda_i \lambda_j \right) G_{ \lambda_{1} \cdots \lambda_{n}} \right | _{ \lambda_j = \frac {1} { \lambda_i}} \notag 
 \\
& \qquad = \sum_{s=0} ^{n-2} \sum _{ \left( j_1, \dots,j_s \right) \in \varTheta_{n \setminus i \setminus j} ^s} \left[ 
 c^{(n)} _{ \left(i,j, j_1 \dots j_s \right) } \Upsilon _{ \left( j_1 \dots j_s \right)} ^n \left ( 1- \lambda_i \lambda_j \right) \Phi _{ \left(i,j, j_1 \dots j_s \right) } 
 \Phi _{ n \setminus i \setminus j \setminus \left( j_1 \dots j_s \right) } \right. \notag
 \\
& \qquad \quad \left. + c^{(n)} _{ \left(j_1 \dots j_s \right) } \Upsilon _{ \left( j_1 \dots j_s \right)} ^n \Phi _{ \left( j_1 \dots j_s \right) } 
 \left ( 1- \lambda_i \lambda_j \right) \Phi _{ n \setminus \left( j_1 \dots j_s \right) } \right ] _{ \lambda_j = \frac {1} { \lambda_i}} \notag 
 \\
& \qquad = \sum_{s=0} ^{n-2} \sum _{ \left( j_1, \dots,j_s \right) \in \varTheta_{n \setminus i \setminus j} ^s} \Upsilon _{ \left( j_1 \dots j_s \right)} ^n \frac{1- \lambda_{i}}{1+ \lambda_{i}} \left[ 
(-1)^{(n+1)(s+2)-i-j- \sum_{k=1}^{s}j_{k}} \times ( -1)^h \right. \notag
 \\
& \qquad \quad \left. + (-1)^{(n+1)s- \sum_{k=1}^{s}j_{k}} \times 
 \left ( -1 \right) ^{j-i-h+1} \right ] \Phi _{ \left(j_1 \dots j_s \right) } \Phi _{ n \setminus i \setminus j \setminus \left( j_1 \dots j_s \right) } \notag \\ 
 & \qquad =0
 \label {eq:e220264}
 \end{align}
where $h$ is as in Eq. \eqref {eq:e220262}. By the above deduction, we can see, for any $ j >i$, $1- \lambda_i \lambda_j$ must be the factors in the denominator polynomial of rational function $G_{ \lambda_{1} \cdots \lambda_{n}}$.

Hence, by Eqs. \eqref{eq:e220262}, \eqref{eq:e220263}and \eqref {eq:e220264}, Eq. \eqref {eq:e220255}is true, and then Eq. \eqref {eq:e220243} holds when $N=n$.
 
 (3) It is assumed that when $N=k$, Eq. \eqref {eq:e220243} holds for any $n$, that is, we have
 \begin{align}
 V_{k}^{ \lambda_{1} \cdots \lambda_{n}}
 = \sum_{s=0} ^n \sum _{ \left( j_1, \dots,j_s \right) \in \varTheta_n ^s} 
 c^{(n)} _{ \left( j_1 \dots j_s \right) } \Upsilon _{ \left( j_1 \dots j_s \right)} ^k \Phi _{ \left( j_1 \dots j_s \right) } 
 \Phi _{ n \setminus \left( j_1 \dots j_s \right) } 
 \label{eq:e220265}
 \end{align}

 (4) Next, according to the inductive method, based on the above step (1), (2) and (3), It is needs to prove when $N=k+1$, Eq. \eqref {eq:e220243} holds for any $n$, that is, the following equation will be needs to be proven true.
 \begin{align}
 V_{k+1}^{ \lambda_{1} \cdots \lambda_{n}}
 = \sum_{s=0} ^n \sum _{ \left( j_1, \dots,j_s \right) \in \varTheta_n ^s} 
 c^{(n)} _{ \left( j_1 \dots j_s \right) } \Upsilon _{ \left( j_1 \dots j_s \right)} ^{k+1} \Phi _{ \left( j_1 \dots j_s \right) } 
 \Phi _{ n \setminus \left( j_1 \dots j_s \right) } 
 \label{eq:e220266} 
 \end{align}

 By the definition of $V_{N}^{ \lambda_{1} \cdots \lambda_{n}}$ and $F_{ \lambda_{1} \cdots \lambda_{n}} ^{ i_{1} \dots i_{n}}$, we have 
 \begin{align} 
 V_{k+1}^{ \lambda_{1} \cdots \lambda_{n}}
& = \sum_{(i_{2}, \cdots,i_{n}) \in \Omega_{1, k}^{n-1}}F_{ \lambda_{1} \cdots \lambda_{n}}^{0 i_{2} \cdots i_{n}}+ \sum_{(i_{1}, \cdots,i_{n}) \in \Omega_{1, k}^{n}}F_{ \lambda_{1} \cdots \lambda_{n}}^{i_{1} \cdots i_{n}} \notag 
 \\
& = \sum_{(i_{2}, \cdots,i_{n}) \in \Omega_{1, k}^{n-1}} \sum_{q=1}^{n}(-1)^{1+k}F_{ \lambda_{1} \cdots \lambda_{n} \setminus \lambda_{k}}^{i_{2} \cdots i_{n}} + \Upsilon_n \sum_{(i_{1}, \cdots,i_{n}) \in \Omega_{0, k-1}^{n}}F_{ \lambda_{1} \cdots \lambda_{n}}^{i_{1} \cdots i_{n}} \notag 
 \\
& = \sum_{(i_{2}, \cdots,i_{n}) \in \Omega_{0, k-1}^{n-1}} \sum_{q=1}^{n}(-1)^{1+q} \Upsilon_{n \setminus q} F_{ \lambda_{1} \cdots \lambda_{n} \setminus \lambda_{q}}^{i_{2} \cdots i_{n}} 
 + \Upsilon_n V_{ k}^{ \lambda_{1} \cdots \lambda_{n}} \notag 
 \\
& = \sum_{q=1}^{n}(-1)^{1+q} \Upsilon_{n \setminus q} V_{ k}^{ \lambda_{1} \cdots \lambda_{n} \setminus \lambda_{q}}+ \Upsilon_n V_{ k}^{ \lambda_{1} \cdots \lambda_{n}} \notag
 \\ 
& = \sum_{q=1}^{n}(-1)^{1+q} \Upsilon_{n \setminus q} 
 \sum_{s=0} ^{n -1} \sum _{ \left( j_1, \dots,j_s \right) \in \varTheta_{n \setminus q} ^s} c^{(n-1)} _{ \left( \hat j_1 \dots \hat j_s \right) } \Upsilon _{ \left( j_1 \dots j_s \right)} ^{k} \Phi _{ \left( j_1 \dots j_s \right) } \Phi _{ n \setminus q \setminus \left( j_1 \dots j_s \right) } \notag
 \\
& \quad + \Upsilon_n 
 \sum_{s=0} ^n \sum _{ \left( j_1, \dots,j_s \right) \in \varTheta_n ^s} c^{(n)} _{ \left( j_1 \dots j_s \right) } \Upsilon _{ \left( j_1 \dots j_s \right)} ^{k} \Phi _{ \left( j_1 \dots j_s \right) } \Phi _{ n \setminus \left( j_1 \dots j_s \right) } \notag
 \\ 
& = \sum_{q=1}^{n}(-1)^{1+q} 
 \sum_{s=0} ^{n -1} \sum _{ \left( j_1, \dots,j_s \right) \in \varTheta_{n \setminus q} ^s} c^{(n-1)} _{ \left( \hat j_1 \dots \hat j_s \right) } \Upsilon_{n \setminus q \setminus \left( j_1 \dots j_s \right) } \Upsilon _{ \left( j_1 \dots j_s \right)} ^{k+1} \Phi _{ \left( j_1 \dots j_s \right) } \Phi _{ n \setminus q \setminus \left( j_1 \dots j_s \right) } \notag
 \\
& \quad + 
 \sum_{s=0} ^n \sum _{ \left( j_1, \dots,j_s \right) \in \varTheta_n ^s} c^{(n)} _{ \left( j_1 \dots j_s \right) } \Upsilon_{n \setminus \left( j_1 \dots j_s \right)} \Upsilon _{ \left( j_1 \dots j_s \right)} ^{k+1} \Phi _{ \left( j_1 \dots j_s \right) } \Phi _{ n \setminus \left( j_1 \dots j_s \right) } \label{eq:e220267}
 \end{align} 
where $ \hat j_1 \dots \hat j_s$ is a new ordinal number sequence because that the ordinal number sequence $ \left( j_1, \dots,j_s \right)$ is produced from the space $ \varTheta_{n \setminus q} ^s$, and then, we have
 \begin {align}
 \hat j_k = \left \{ \begin{array}{ll} j_k & j_k<q \\
 j_k-1 & j_k>q
 \end{array}
 \right . \label{eq:e220268}
 \end{align}
 \\
 Comparing between Eqs. \eqref{eq:e220266} and \eqref {eq:e220267}, for proving Eq. \eqref{eq:e220266}, the following equation should be proven at first.
 \begin{align}
 \sum_{q=1}^{n} & (-1)^{1+q}
 \sum_{s=0} ^{n -1} \sum _{ \left( j_1, \dots,j_s \right) \in \varTheta_{n \setminus q} ^s} c^{(n-1)} _{ \left( \hat j_1 \dots \hat j_s \right) } \Upsilon_{n \setminus q \setminus \left( j_1 \dots j_s \right) } \Upsilon _{ \left( j_1 \dots j_s \right)} ^{k+1} \Phi _{ \left( j_1 \dots j_s \right) } \Phi _{ n \setminus q \setminus \left( j_1 \dots j_s \right) } \notag
 \\
 & \quad + 
 \sum_{s=0} ^n \sum _{ \left( j_1, \dots,j_s \right) \in \varTheta_n ^s} c^{(n)} _{ \left( j_1 \dots j_s \right) }
 \left ( \Upsilon_{n \setminus \left( j_1 \dots j_s \right)} -1 \right) \Upsilon _{ \left( j_1 \dots j_s \right)} ^{k+1} \Phi _{ \left( j_1 \dots j_s \right) } \Phi _{ n \setminus \left( j_1 \dots j_s \right) } \notag \\
 & =0 \label{eq:e220269}
 \end{align}

In fact, the left side of the above equation can be classified and rearranged according to the factor $ \Upsilon _{ \left( j_1 \dots j_s \right)} ^{k+1} \Phi _{ \left( j_1 \dots j_s \right) }$ and then, if the sum of the factors accompanying with the factor $ \Upsilon _{ \left( j_1 \dots j_s \right)} ^{k+1} \Phi _{ \left( j_1 \dots j_s \right) }$ can be proven as zero, Eq. \eqref {eq:e220269} can be proved successfully. The proving process can be exampled as 
follows.

1) For the factor $ \Upsilon _{ 1} ^{k+1} \Phi _{ 1 }= \lambda_1 ^{k+1} \Phi _{ \lambda_1 }$, by \textbf{Lemma \ref {le:L22020102} } and Eq. \eqref {eq:e220246}, the sum of the accompanying factors in the left side of Eq. \eqref {eq:e220269} is as follows.
 \begin{align}
 \sum_{q=2}^{n} & (-1)^{1+q}
 c^{(n-1)} _{ 1 } \Upsilon_{n \setminus q \setminus 1} \Phi _{ n \setminus q \setminus 1 } + 
c^{(n)} _{ 1 }
 \left ( \Upsilon_{n \setminus 1} -1 \right) \Phi _{ n \setminus 1 } \notag \\
 & = (-1)^{n-1}
 \sum_{q=2}^{n} (-1)^{1+q} \Upsilon_{n \setminus q \setminus 1} \Phi _{ n \setminus q \setminus 1 } - 
 (-1)^{n} \sum_{q=1}^{n-1} (-1)^{1+q}
 \Upsilon_{n \setminus q+1 \setminus 1} \Phi _{ n \setminus q+1 \setminus 1 } \notag \\
 & = 
 \sum_{q=2}^{n} (-1)^{n+q} \Upsilon_{n \setminus q \setminus 1} \Phi _{ n \setminus q \setminus 1 } - 
 \sum_{q=2}^{n} (-1)^{n+q}
 \Upsilon_{n \setminus q \setminus 1} \Phi _{ n \setminus q \setminus 1 } \notag \\
 & =0 \label{eq:e220270}
 \end{align}

2) For the factor $ \Upsilon _{ \left( j_1 \dots j_s \right)} ^{k+1} \Phi _{ \left( j_1 \dots j_s \right) }$, by \textbf{Lemma \ref {le:L22020102} } and Eq. \eqref {eq:e220246}, the sum of the accompanying factors in the left side of Eq. \eqref {eq:e220269} is as follows.
 \begin{align}
 \sum_{q=1}^{n \setminus \left( j_1 \dots j_s \right) } & (-1)^{1+q}
 c^{(n-1)} _{ \left( n_{j_1} \dots n_{j_s} \right) } \Upsilon_{n \setminus q \setminus \left( j_1 \dots j_s \right) } \Phi _{ n \setminus q \setminus \left( j_1 \dots j_s \right) } \notag \\
 & \qquad \quad +
 c^{(n)} _{ \left( j_1 \dots j_s \right) }
 \left ( \Upsilon_{n \setminus \left( j_1 \dots j_s \right)} -1 \right) \Phi _{ n \setminus \left( j_1 \dots j_s \right) } \notag \\
& = \sum_{q=1}^{n \setminus \left( j_1 \dots j_s \right)} (-1)^{1+q}
c^{(n-1)} _{ \left( n_{j_1} \dots n_{j_s} \right) } \Upsilon_{n \setminus q \setminus \left( j_1 \dots j_s \right) } \Phi _{ n \setminus q \setminus \left( j_1 \dots j_s \right) } \notag \\
& \qquad \quad -
c^{(n)} _{ \left( j_1 \dots j_s \right) } \sum_{q=1}^{n \setminus \left( j_1 \dots j_s \right) } (-1)^{1+ \hat q}
 \Upsilon_{n \setminus q \setminus \left( j_1 \dots j_s \right) } \Phi _{ n \setminus q \setminus \left( j_1 \dots j_s \right) } \notag \\
& = \sum_{q=1}^{n \setminus \left( j_1 \dots j_s \right)} \left[ (-1)^{1+q}
c^{(n-1)} _{ \left( \hat j_1 \dots \hat j_s \right) } -
c^{(n)} _{ \left( j_1 \dots j_s \right) } (-1)^{1+ \hat q} \right]
 \Upsilon_{n \setminus q \setminus \left( j_1 \dots j_s \right) } \Phi _{ n \setminus q \setminus \left( j_1 \dots j_s \right) } \label{eq:e220271}
 \end{align}
 where $ \hat q$ is the position number of the number '$q$' in the ordinal number sequence $(1 2 \dots n) \setminus \left( j_1 \dots j_s \right)$,
 that is, we have
 \begin{align} 
 \hat q = 
 q-k, \; \; j_{k}< q < j_{k+1} \label {eq:e220272} 
 \end{align} 
In addition, by Eq. \eqref {eq:e220268}, we have
 \begin{align} 
 \sum _{k=1} ^s \hat j_k= \sum _{k=1} ^s j_k -(s-k), \; \; j_{k}< q < j_{k+1}
 \label {eq:e220273} 
 \end{align} 
 
 Therefore, by Eqs. \eqref {eq:e220272} and \eqref {eq:e220273}, if $ j_{k}< q < j_{k+1}$, we have 
 \begin{align} 
(-1)^{1+q} & c^{(n-1)} _{ \left( \hat j_1 \dots \hat j_s \right) } -
c^{(n)} _{ \left( j_1 \dots j_s \right) } (-1)^{1+ \hat q}
 \notag \\
& =(-1)^{ns- \sum _{k=1} ^s \hat j_k+1+q}-(-1)^{(n+1)s- \sum _{k=1} ^s j_k+1+ \hat q}
 \notag \\
& =(-1)^{ns- \sum _{k=1} ^s j_k-(s-k)+1+q}-(-1)^{(n+1)s- \sum _{k=1} ^s j_k+1+ q-k}
 \notag \\
& =0
 \label {eq:e220274} 
 \end{align} 

In summary, by Eqs. \eqref {eq:e220271} and \eqref {eq:e220274}, Eqs. \eqref{eq:e220269} and \eqref {eq:e220266} are proved successively as true. 

Synthesized the step (1) to (4) in the inductive process, the theorem has been proven. 
 \qed

Based on the above theorem, for $n=2,3$, we have 
 \begin{align}
 V_N^{ \lambda_1 \lambda_2}
 & = \Phi_{ \lambda_{1} \lambda_{2}} + \left( \lambda_{1}^{N}- \lambda_{2}^{N} \right) \Phi_{ \lambda_{1} } \Phi_{ \lambda_{2} } - \lambda_{1}^{N} \lambda_{2}^{N} \Phi_{ \lambda_{1} \lambda_{2}}
 \notag \\
 & = \frac{ \left ( \lambda_{2}- \lambda_{1} \right) \left(1 - \lambda_{1}^{N} \lambda_{2}^{N} \right) }{ \left(1- \lambda_{1} \right) \left(1- \lambda_{2} \right) \left(1- \lambda_{2} \lambda_{1} \right)}+ \frac{ \lambda_{1}^{N}- \lambda_{2}^{N}}{ \left(1- \lambda_{1} \right) \left(1- \lambda_{2} \right)} \label{eq:e220275}
 \\
 V_N^{ \lambda_1 \lambda_2 \lambda_3}
 & = \Phi_{ \lambda_{1} \lambda_{2} \lambda_3} - \lambda_{1}^{N} \Phi_{ \lambda_{1} } \Phi_{ \lambda_{2} \lambda_{3} } + \lambda_{2}^{N} \Phi_{ \lambda_{2} } \Phi_{ \lambda_{1} \lambda_{3} } - \lambda_{3}^{N} \Phi_{ \lambda_{3} } \Phi_{ \lambda_{1} \lambda_{2} } \notag \\
 & \quad - \lambda_{1}^{N} \lambda_{2}^{N} \Phi_{ \lambda_{1} \lambda_{2} } \Phi_{ \lambda_{3} } + \lambda_{1}^{N} \lambda_{3}^{N} \Phi_{ \lambda_{1} \lambda_{3} } \Phi_{ \lambda_{2} } - \lambda_{2}^{N} \lambda_{3}^{N} \Phi_{ \lambda_{2} \lambda_{3} } \Phi_{ \lambda_{1} }
 \notag \\
 & \quad + \lambda_{1}^{N} \lambda_{2}^{N} \lambda_{3}^{N} \Phi_{ \lambda_{1} \lambda_{2} \lambda_3 }
 \notag \\
 & = \left(1 + \lambda_{1}^{N} \lambda_{2}^{N} \lambda_{3}^{N} \right) \Phi_{ \lambda_{1} \lambda_{2} \lambda_3} 
 - \left( \lambda_{1}^{N} + \lambda_{2}^{N} \lambda_{3}^{N} \right) \Phi_{ \lambda_{1} } \Phi_{ \lambda_{2} \lambda_3}
 \notag \\
 & \quad + \left( \lambda_{2}^{N} + \lambda_{1}^{N} \lambda_{3}^{N} \right) \Phi_{ \lambda_{2} } \Phi_{ \lambda_{1} \lambda_3}
 - \left( \lambda_{3}^{N} + \lambda_{1}^{N} \lambda_{2}^{N} \right) \Phi_{ \lambda_{3} } \Phi_{ \lambda_{1} \lambda_2} \label{eq:e220276}
 \end{align}

 \section{The analytical factors describing the finite-time control capability }

 By \textbf{Theorem \ref{th:t22020104}}, the factor with the biggest absolute value in the sum expression \eqref {eq:e220242} and \eqref{eq:e220243} is as follows
 \begin{align}
 \left \vert \det \left ( W^{-1} \right) \prod _{i=1} ^n q_ib \right \vert 
 \times 
 \left \vert \Upsilon _{ \left( j_1 \dots j_s \right)} ^N \Phi _{ \left( j_1 \dots j_s \right) } 
 \Phi _{ n \setminus \left( j_1 \dots j_s \right) } \right \vert
 \label{eq:e220277} \end{align}
where $ \left \{ \lambda _{j_1}, \lambda_{j_2}, \dots, \lambda_{j_s} \right \}$ are all eigenvalues of matrix $A$ satisfying $ \lambda_{j_i}>1$. Therefore,
some analytical factors describing the finite-time control capability of the LDTs can be extracted from the analytical expression of the volume of the controllable region as follows
 \begin{align} 
 F_1 & =F_{1}^{+} \times F_{1}^{-} \notag \\
 &= \left \vert \prod_{ \forall j_{1},j_{2} \in P^+} \frac{ \lambda_{j_{2}}- \lambda_{j_{1}} }{ 1- \lambda_{j_{1}} \lambda_{j_{2}}} \right \vert \times 
 \left \vert \prod_{ \forall k_{1},k_{2} \in P^-} \frac{ \lambda_{k_{2}}- \lambda_{k_{1}} }{ 1- \lambda_{k_{1}} \lambda_{k_{2}}} \right \vert
 \label{eq:e220278} \\
 F_{2,i} 
 &= \frac{ \left \vert q_ib \right \vert \left( 1- \lambda_{i} ^N \right) \vert}{ 1- \lambda_{i} }, \; \; i=1,2, \dots,n
 \label{eq:e220279} \\
 F_{3,i} & = \left \vert q_ib \right \vert, \; \;i=1,2, \dots, n \label{eq:e220280} 
 \end{align}
where $P^+$ and $P^-$ are defined as follows
 \begin{align}
 P^+ & = \left \{ \lambda_i: \lambda_i>1, \forall i=1,2, \dots, n \right \} \notag \\
 P^- & = \left \{ \lambda_i: \lambda_i<1, \forall i=1,2, \dots, n \right \} \notag
 \end{align}
As discussing in papers \cite {zhaomw202002}, \cite {zhaomw202003} and \cite {zhaomw202004}, the above analytical factors can be called respectively as 
the shape(pole distribution factor) factor, the side length of the circumscribed rhombohedral, and the modal controllability, and these factors are with same significances to these papers. 

 \section {Three generalized cases of Theorem \ref{th:t22020104} }

 \subsection{The Narrow control capability and its analytical computing}

Strictly speaking, as discussing above the control capability discused above is a broad control capability (also called reachable capability). Based on Eq. \eqref {eq:e220207} to Eq. \eqref {eq:e220210}, The above analytical analysis results on the reachable region$R_N^d(A)$ can be generlized to the narrow controllable region $R_N^c(A)$. 
Therefor, based on the above relations and \textbf{Theorem \ref{th:t22020104}}, we have the following theorem about the analytical computation for the volume of the region $R_N^c(A)$.
 
 \begin{theorem} \label{th:t22020105}
 If the $n$ eigenvalues $ \lambda_i(i= \overline{1,n})$ of the matrix $A$ are positive real numbers and satisfy 
 $$
 0< \lambda_1< \lambda_2< \cdots< \lambda_n,
 $$
 the volume of the finite-time narrow controllable region $R^c _N (A)$ when $N \ge n$ can be computed analytical as follows
 \begin{align}
 V_n \left(R^c _N (A) \right) &=2^n \left \vert \det \left ( W^{-1} \right) \prod _{i=1} ^n \beta_i \right \vert V_{N}^{ \lambda_{1} \cdots \lambda_{n}} \label{eq:e220281}\\
 V_{N}^{ \lambda_{1} \cdots \lambda_{n}}
 & = \sum_{s=0} ^n \sum _{ \left( j_1, \dots,j_s \right) \in \varTheta_n ^s} 
 c^{(n)} _{ \left( j_1 \dots j_s \right) } \Upsilon _{ \left( j_1 \dots j_s \right)} ^{-N} \Phi _{ \left( j_1 \dots j_s \right) } 
 \Phi _{ n \setminus \left( j_1 \dots j_s \right) } 
 \label{eq:e220282} \end{align}
 where 
 \begin{align}
 \Upsilon _{ \left( j_1 \dots j_s \right)} ^{-N} = \prod_{k=1} ^{ s} \lambda _{j_k} ^{-N} \label{eq:e220283}
 \end{align} 
 \end{theorem}

By \textbf{Theorem \ref{th:t22020105}}, the factor with the biggest absolute in the sum expression \eqref {eq:e220281}and \eqref {eq:e220282} is as follows
 \begin{align}
 \left \vert \det \left ( W^{-1} \right) \prod _{i=1} ^n q_ib \right \vert 
 \times 
 \left \vert \Upsilon _{ \left( j_1 \dots j_s \right)} ^{-N} \Phi _{ \left( j_1 \dots j_s \right) } 
 \Phi _{ n \setminus \left( j_1 \dots j_s \right) } \right \vert
 \label{eq:e220284} \end{align}
where $ \left \{ \lambda _{j_1}, \lambda_{j_2}, \dots, \lambda_{j_s} \right \}$ are all eigenvalues of matrix $A$ satisfying $ \lambda_{j_i}<1$. Therefore,
some analytical factors describing the finite-time control capability of the LDTs can be extracted from the analytical expression of the volume of the controllable region as follows
 \begin{align} 
 F_1 & =F_{1}^{+} \times F_{1}^{-} \notag \\
 &= \left \vert \prod_{ \forall j_{1},j_{2} \in P^+} \frac{ \lambda_{j_{2}}- \lambda_{j_{1}} }{ 1- \lambda_{j_{1}} \lambda_{j_{2}}} \right \vert \times 
 \left \vert \prod_{ \forall k_{1},k_{2} \in P^-} \frac{ \lambda_{k_{2}}- \lambda_{k_{1}} }{ 1- \lambda_{k_{1}} \lambda_{k_{2}}} \right \vert
 \label{eq:e220285} \\
 F_{2,i} 
 &= \frac{ \left \vert q_ib \right \vert \left( 1- \lambda_{i} ^{-N} \right) \vert}{ 1- \lambda_{i} }, \; \;i=1,2, \dots,n
 \label{eq:e220286} \\
 F_{3,i} & = \left \vert q_ib \right \vert, \; \;i=1,2, \dots,n \label{eq:e220287} 
 \end{align}
where $P^+$ and $P^-$ are defined as follows
 \begin{align}
 P^+ & = \left \{ \lambda_i: \lambda_i>1, \forall i=1,2, \dots, n \right \} \notag \\
 P^- & = \left \{ \lambda_i: \lambda_i<1, \forall i=1,2, \dots, n \right \} \notag
 \end{align}
The above analytical factors can be called respectively as 
the shape(pole distribution factor) factor, the side length of the circumscribed rhombohedral, and the modal controllability. In fact, the shape factor $F_1$ is also the eigenvalue evenness factor of the linear system, and can describe the control capability caused by the eigenvalue distribution. In addition, the modal controllability factor $F_{3,i}$ have been put forth by papers \cite{chan1984} \cite{HamElad1988} \cite{ChoParLee2000} \cite{Chenliu2001}, and will not be discussed here.

 \subsection {All eigenvalues of the matrix $A$ are negative }

 \textbf{Theorem \ref {th:t22020104}} can be generalized the case that all eigenvalues of the natrix $A$ are negative as follows.
 \begin{theorem} \label{th:t22020106}
 If the $n$ eigenvalues $ \lambda_i(i= \overline{1,n})$ of the matrix $A$ are negative real numbers and satisfy 
 $$
 \lambda_1< \lambda_2< \cdots< \lambda_n<0,
 $$
 the volume of the finite-time zonotope $E_n (P_N)$ spaned by the columns of the matrix $P_N= \left[b,Ab, \dots , A^{N-1}b \right]$ when $N \ge n$ can be computed analytical as follows
 \begin{align}
 V_n \left(E_n \left(P_N \right) \right) &=2^n \left \vert \det \left ( W^{-1} \right) \prod _{i=1} ^n \beta_i \right \vert V_{N}^{ \lambda_{1} \cdots \lambda_{n}} \label{eq:e220288} \\
 V_{N}^{ \lambda_{1} \cdots \lambda_{n}}
 & = \sum_{s=0} ^n \sum _{ \left( j_1, \dots,j_s \right) \in \varTheta_n ^s} 
 c^{(n)} _{ \left( j_1 \dots j_s \right) } \Upsilon _{ \left( j_1 \dots j_s \right)} ^{N} \Phi _{ \left( j_1 \dots j_s \right) } 
 \Phi _{ n \setminus \left( j_1 \dots j_s \right) } 
 \label{eq:e220289} \end{align}
 where 
 \begin{align}
& \Upsilon _{ \left( j_1 \dots j_s \right)} ^{N} = \prod_{k=1} ^{ s} \left \vert \lambda _{j_k} \right \vert ^{N} \label{eq:e220290} \\
 & \Phi _{ \left( j_1 \dots j_s \right) } = \left \vert \prod_{1 \leq i<k \leq s} \frac{ \lambda_{j_{k}}- \lambda_{j_{i}}}{1- \lambda_{j_{i}} \lambda_{j_{k}}} \right \vert \left( \prod_{i=1}^{s} \frac{1}{1+ \lambda_{j_i}} \right) \label{eq:e220291}
 \end{align} 
 \end{theorem}

 \subsection {The volume computing of the controllable region of the linear continuous-time systems }

As discussing in the paper \cite {zhaomw202101}, the volume computing of the controllable regions of the LDT systems can be generalized to the linear continuous-time (LCT) systems $ \Sigma \left ( A_c,B_c \right)$, and then we have the correspongding theorem to \textbf{Theorem \ref {th:t22020104}} as follows
 \begin{theorem} \label{th:t22020107}
 If the $n$ eigenvalues $ \left ( \lambda_i, i= \overline{1,n} \right ) $ of the matrix $A$ are real numbers and the smothing zonotope $R_T \left ( A_c \right) $ generated by the matrix pair $ \Sigma \left ( A_c,B_c \right)$ in the finite time $[0,T]$ is defined as follows
 \begin{align}
 R_T \left ( A_c \right) = \left \{ x:x= \int _0^T \exp \left(A_c t \right) B_c u_t \textnormal {d}t, \forall u_t \in [-1,1] \right \}  \label{eq:e220292}
 \end{align}
the volume of $R_T \left ( A_c \right)$ in the finite-time $T$ can be computed analytical as follows
 \begin{align}
 V_n \left(R_T \left ( A_c \right) \right) &=2^n \left \vert \det \left ( W^{-1} \right) \prod _{i=1} ^n \beta_i \right \vert V_{T}^{ \lambda_{1} \cdots \lambda_{n}} \label{eq:e220293} \\
 V_{T}^{ \lambda_{1} \cdots \lambda_{n}}
 & = \sum_{s=0} ^n \sum _{ \left( j_1, \dots,j_s \right) \in \varTheta_n ^s} 
 c^{(n)} _{ \left( j_1 \dots j_s \right) } \Upsilon _{ \left( j_1 \dots j_s \right)} ^{T} \Phi _{ \left( j_1 \dots j_s \right) } 
 \Phi _{ n \setminus \left( j_1 \dots j_s \right) } 
 \label{eq:e220294} \end{align}
 where 
 \begin{align}
 & \Upsilon _{ \left( j_1 \dots j_s \right)} ^{T} = \exp \left ( \sum_{k=1} ^{ s} \lambda _{j_k} T \right ) \label{eq:e220295} \\
 & \Phi _{ \left( j_1 \dots j_s \right) } = \left \vert \prod_{1 \leq i<k \leq s} \frac{ \lambda_{j_{k}}- \lambda_{j_{i}}}{ \lambda_{j_{i}} + \lambda_{j_{k}}} \right \vert \left( \prod_{i=1}^{s} \frac{1}{ \lambda_{j_i}} \right) \label{eq:e220296}
 \end{align} 
 \end{theorem}

 By the above theorem, the factor with the biggest absolute in the sum expression \eqref {eq:e220293} and \eqref{eq:e220294} is as follows
 \begin{align}
 \left \vert \det \left ( W^{-1} \right) \prod _{i=1} ^n q_ib \right \vert 
 \times 
 \left \vert \Upsilon _{ \left( j_1 \dots j_s \right)} ^T \Phi _{ \left( j_1 \dots j_s \right) } 
 \Phi _{ n \setminus \left( j_1 \dots j_s \right) } \right \vert
 \label{eq:e220297} \end{align}
where $ \left \{ \lambda _{j_1}, \lambda_{j_2}, \dots, \lambda_{j_s} \right \}$ are all eigenvalues of matrix $A$ satisfying $ \lambda_{j_i}>0$. Therefore,
some analytical factors describing the finite-time control capability of the LDTs can be extracted from the analytical expression of the volume of the controllable region as follows
 \begin{align} 
 F_1 & =F_{1}^{+} \times F_{1}^{-} \notag \\
 &= \left \vert \prod_{ \forall j_{1},j_{2} \in P^+} \frac{ \lambda_{j_{2}}- \lambda_{j_{1}} }{ \lambda_{j_{1}} + \lambda_{j_{2}}} \right \vert \times 
 \left \vert \prod_{ \forall k_{1},k_{2} \in P^-} \frac{ \lambda_{k_{2}}- \lambda_{k_{1}} }{ \lambda_{k_{1}} + \lambda_{k_{2}}} \right \vert
 \label{eq:e220298} \\
 F_{2,i} 
 &= \frac{ \left \vert q_ib \right \vert \left \vert 1- \exp \left ( \lambda_{i} T \right ) \right \vert}{ \lambda_{i} }, \; \; i=1,2, \dots,n
 \label{eq:e220299} \\
 F_{3,i} & = \left \vert q_ib \right \vert, \; \;i=1,2, \dots, n \label{eq:e2202A00} 
 \end{align}
where $P^+$ and $P^-$ are defined as follows
 \begin{align}
 P^+ & = \left \{ \lambda_i: \lambda_i>0, \forall i=1,2, \dots, n \right \} \notag \\
 P^- & = \left \{ \lambda_i: \lambda_i<0, \forall i=1,2, \dots, n \right \} \notag
 \end{align}
As discussing in papers \cite {zhaomw202002}, \cite {zhaomw202003} and \cite {zhaomw202004}, the above analytical factors can be called respectively as 
the shape(pole distribution factor) factor, the side length of the circumscribed rhombohedral, and the modal controllability, and these factors are with same significances to these papers. 
 \\
 \section {Numerical Experiments (Not available here)}

 \section{Conclusions (Not available here)}

 \bibliographystyle{model1b-num-names}
 \bibliography{zzz}

\begin{thebibliography}{19}
\expandafter\ifx\csname natexlab\endcsname\relax\def\natexlab#1{#1}\fi
\providecommand{\bibinfo}[2]{#2}
\ifx\xfnm\relax \def\xfnm[#1]{\unskip,\space#1}\fi
%Type = Inbook
\bibitem[{Beck and Robins(2015)}]{BecRob2015}
\bibinfo{author}{M.~Beck}, \bibinfo{author}{S.~Robins},
  \bibinfo{title}{Computing the continuous discretely: Integer-point
  enumeration in polyhedra}, \bibinfo{title}{Computing the Continuous
  Discretely: Integer-Point Enumeration in Polyhedra},
  \bibinfo{publisher}{Springer}, \bibinfo{year}{2015}, \bibinfo{edition}{2nd
  edition} edition, pp. \bibinfo{pages}{167--182}.
%Type = Article
\bibitem[{Chan(1984)}]{chan1984}
\bibinfo{author}{S.~Chan}, \bibinfo{title}{Modal controllability and
  observability of power-system models}, \bibinfo{journal}{International
  Journal of Electrical Power \& Energy Systems} \bibinfo{volume}{6}
  (\bibinfo{year}{1984}) \bibinfo{pages}{83--88}.
%Type = Book
\bibitem[{Chen(1998)}]{Chen1998}
\bibinfo{author}{C.T. Chen}, \bibinfo{title}{Linear system theory and design},
  \bibinfo{publisher}{Oxford University Press, Inc. New York, NY, USA},
  \bibinfo{edition}{3rd} edition, \bibinfo{year}{1998}.
%Type = Article
\bibitem[{Chen et~al.(2001)Chen, Chen and Liu}]{Chenliu2001}
\bibinfo{author}{Y.~Chen}, \bibinfo{author}{S.~Chen}, \bibinfo{author}{Z.~Liu},
  \bibinfo{title}{Quantitative measures of modal controllability and
  observability in vibration control of defective and near-defective systems},
  \bibinfo{journal}{Journal of Sound and Vibration} \bibinfo{volume}{248}
  (\bibinfo{year}{2001}) \bibinfo{pages}{413--26}.
%Type = Inproceedings
\bibitem[{Choi et~al.(2000)Choi, Park and Lee}]{ChoParLee2000}
\bibinfo{author}{J.W. Choi}, \bibinfo{author}{U.S. Park}, \bibinfo{author}{S.B.
  Lee}, \bibinfo{title}{Measures of modal controllability and observability in
  balanced coordinates for optimal placement of sensors and actuators: a
  flexible structure application}, in: \bibinfo{booktitle}{Proceedings of the
  SPIE}, volume \bibinfo{volume}{3984}, pp. \bibinfo{pages}{425--436}.
%Type = Inproceedings
\bibitem[{Georges(1995)}]{Georges1995}
\bibinfo{author}{D.~Georges}, \bibinfo{title}{The use of observability and
  controllability gramians or functions for optimal sensor and actuator
  location in finite-dimensional systems}, in: \bibinfo{booktitle}{Proc. of
  IEEE Conf. on Decision and Control}, \bibinfo{address}{New Orleans, LA, USA},
  p. \bibinfo{pages}{3319–3324}.
%Type = Article
\bibitem[{Gover and Krikorian(2010)}]{GovKri2010}
\bibinfo{author}{E.~Gover}, \bibinfo{author}{N.~Krikorian},
  \bibinfo{title}{Determinants and the volumes of parallelotopes and
  zonotopes}, \bibinfo{journal}{Linear Algebra and its Applications 433 (2010)
  28–40} \bibinfo{volume}{433} (\bibinfo{year}{2010})
  \bibinfo{pages}{28–40}.
%Type = Article
\bibitem[{Hamdan and Eladbdalla(1988)}]{HamElad1988}
\bibinfo{author}{A.~Hamdan}, \bibinfo{author}{A.~Eladbdalla},
  \bibinfo{title}{Geometric measures of modal controllability and observability
  of power system models}, \bibinfo{journal}{Electric Power Systems Research}
  \bibinfo{volume}{15} (\bibinfo{year}{1988}) \bibinfo{pages}{147--155}.
%Type = Phdthesis
\bibitem[{Ilkturk(2015)}]{Ilkturk2015}
\bibinfo{author}{U.~Ilkturk}, \bibinfo{title}{Observability Methods in Sensor
  Scheduling}, Ph.D. thesis, ARIZONA STATE UNIVERSITY, \bibinfo{year}{2015}.
%Type = Book
\bibitem[{Kailath(1980)}]{Kailath1980}
\bibinfo{author}{T.~Kailath}, \bibinfo{title}{Linear systems},
  \bibinfo{publisher}{Prentice-Hall}, \bibinfo{address}{Englewood Cliffs, NJ},
  \bibinfo{year}{1980}.
%Type = Article
\bibitem[{Kalman et~al.(1963)Kalman, Ho and Narendra}]{KalmHoNar1963}
\bibinfo{author}{R.E. Kalman}, \bibinfo{author}{Y.C. Ho}, \bibinfo{author}{K.S.
  Narendra}, \bibinfo{title}{Controllability of dynamical systems},
  \bibinfo{journal}{Contributions to Differential Equations}
  \bibinfo{volume}{1} (\bibinfo{year}{1963}) \bibinfo{pages}{189--213}.
%Type = Article
\bibitem[{McMullen(????)}]{McMu1971}
\bibinfo{author}{P.~McMullen}, \bibinfo{title}{On zonotopes},
  \bibinfo{journal}{Trans. of the Aerican Mathematical Society}
  \bibinfo{volume}{159} (????) \bibinfo{pages}{91--109}.
%Type = Article
\bibitem[{Pasqualetti et~al.(2014)Pasqualetti, Zampieri and
  Bullo}]{PasaZamEul2014}
\bibinfo{author}{F.~Pasqualetti}, \bibinfo{author}{S.~Zampieri},
  \bibinfo{author}{F.~Bullo}, \bibinfo{title}{Controllability metrics,
  limitations and algorithms for complex networks}, \bibinfo{journal}{IEEE
  Trans. on Control of Network Systems} \bibinfo{volume}{1}
  (\bibinfo{year}{2014}) \bibinfo{pages}{40--52}.
%Type = Techreport
\bibitem[{VanderVelde and Carignan(1982)}]{VanCari1982}
\bibinfo{author}{W.~VanderVelde}, \bibinfo{author}{C.~Carignan},
  \bibinfo{title}{A dynamic measure of controllability and observability for
  the placement of actuators and sensors on large space structures},
  \bibinfo{type}{Technical Report}, NASA-CR-168520, SSL-2-82,
  \bibinfo{year}{1982}.
%Type = Article
\bibitem[{Zhao(2020{\natexlab{a}})}]{zhaomw202002}
\bibinfo{author}{M.~Zhao}, \bibinfo{title}{Analytical expression and
  deconstruction of the volume of the controllability ellipsoid},
  \bibinfo{journal}{arXiv:2004.05528}  (\bibinfo{year}{2020}{\natexlab{a}})
  \bibinfo{pages}{10}.
%Type = Article
\bibitem[{Zhao(2020{\natexlab{b}})}]{zhaomw202004}
\bibinfo{author}{M.~Zhao}, \bibinfo{title}{Analytical factors for describing
  the control ability of linear discrete-time systems},
  \bibinfo{journal}{arXiv:2004.07982}  (\bibinfo{year}{2020}{\natexlab{b}}).
%Type = Article
\bibitem[{Zhao(2020{\natexlab{c}})}]{zhaomw202001}
\bibinfo{author}{M.~Zhao}, \bibinfo{title}{Exact volume of zonotopes generated
  by a matrix pair}, \bibinfo{journal}{arXiv:2004.05530}
  (\bibinfo{year}{2020}{\natexlab{c}}) \bibinfo{pages}{20}.
%Type = Article
\bibitem[{Zhao(2020{\natexlab{d}})}]{zhaomw202003}
\bibinfo{author}{M.~Zhao}, \bibinfo{title}{Relations among open-loop control
  ability, control strategy space and closed-loop performance for linear
  discrte-time systems}, \bibinfo{journal}{arXiv:2004.05619}
  (\bibinfo{year}{2020}{\natexlab{d}}) \bibinfo{pages}{13}.
%Type = Article
\bibitem[{Zhao(2021)}]{zhaomw202101}
\bibinfo{author}{M.~Zhao}, \bibinfo{title}{Control capability with time
  attributy for linear continuous-time systems},
  \bibinfo{journal}{arXiv:2103.15038}  (\bibinfo{year}{2021}).

\end{thebibliography}
 \end{document}